\pdfoutput=1
\pdfoutput=1
\documentclass[conference]{IEEEtran}

\usepackage{graphicx}
\usepackage{booktabs}
\usepackage{amsmath}
\usepackage{amssymb}
\usepackage{xcolor}
\usepackage{url}
\usepackage{xurl}
\usepackage{array}
\usepackage{tabularx}
\usepackage{enumitem}
\usepackage{xstring}
\usepackage{tikz}
\usetikzlibrary{fit,positioning,backgrounds,calc}
\definecolor{fgZone}{HTML}{E6EAE4}\definecolor{fgZoneLn}{HTML}{BCC4BF}\definecolor{fgInner}{HTML}{F4F6F2}
\definecolor{fgBox}{HTML}{FBFCFA}\definecolor{fgBoxLn}{HTML}{8E9893}\definecolor{fgInk}{HTML}{1A221F}
\definecolor{fgSub}{HTML}{5D6A64}\definecolor{fgKey}{HTML}{B0284F}\definecolor{fgNew}{HTML}{A86608}
\definecolor{fgBlue}{HTML}{2A62A8}\definecolor{fgGreen}{HTML}{2C7A47}\definecolor{fgArrow}{HTML}{39443F}
\tikzset{
  fg/.style={font=\sffamily\scriptsize, text=fgInk, >=stealth, line cap=round},
  fgzone/.style={draw=fgZoneLn, fill=fgZone, rounded corners=3pt, line width=0.4pt, inner sep=3.5pt},
  fginner/.style={draw=fgZoneLn, fill=fgInner, rounded corners=2.5pt, line width=0.4pt, inner sep=3pt},
  fgbox/.style={draw=fgBoxLn, fill=fgBox, rounded corners=1.6pt, line width=0.5pt, align=center, inner sep=1.6pt, minimum height=4.8mm},
  fgnewbox/.style={fgbox, draw=fgNew, fill=fgNew!13, line width=0.6pt},
  fgkeybox/.style={fgbox, draw=fgKey, line width=0.9pt},
  fgeb/.style={font=\sffamily\tiny\bfseries, text=fgSub, inner sep=0.5pt},
  fgsub/.style={font=\sffamily\tiny, text=fgSub},
  fgar/.style={draw=fgArrow, line width=0.55pt},
  fgbound/.style={draw=fgKey, line width=0.6pt, dash pattern=on 3pt off 2pt},
  fgmk/.style={draw=fgKey, text=fgKey, circle, inner sep=0.35pt, font=\sffamily\tiny\bfseries, fill=white, line width=0.5pt},
}
\newcommand{\fgS}[1]{{\tiny\color{fgSub}#1}}
\newcommand{\fgT}[2]{\textbf{#1}\\[-1.5pt]\fgS{#2}}

\makeatletter\providecommand{\tikzscope@linewidth}{0.4pt}\makeatother

\usepackage[hidelinks]{hyperref}
\hypersetup{pdfauthor={Jungmin Park, Eunha Kim, Wooseop Kim, Seongjoon Cho, Byungho Cha},
 pdftitle={From FIPS to FPGA with LLM Agents: Measuring What Tests and Reviews Miss in Post-Quantum Cryptographic Hardware}}

\newcommand{\cTwoN}{129}
\newcommand{\cTwoSucc}{126}
\newcommand{\cTwoSuccPct}{97.7}
\newcommand{\cTwoWilsonLo}{93}
\newcommand{\cTwoWilsonHi}{99}
\newcommand{\cTwoOneShot}{43}
\newcommand{\cTwoOneShotPct}{33.3}
\newcommand{\cTwoEditFree}{106}
\newcommand{\cTwoEditFreePct}{82.2}
\newcommand{\cTwoIterMax}{7}
\newcommand{\cTwoOracleHw}{37}
\newcommand{\cTwoOracleMachine}{87}
\newcommand{\cTwoOracleMachinePct}{67.4}
\newcommand{\cTwoModelFableFiveN}{33}
\newcommand{\cTwoModelFableFiveOneN}{24}
\newcommand{\cTwoModelOpusFiveFiveN}{66}
\newcommand{\cTwoModelOpusFiveN}{2}
\newcommand{\cTwoModelOpusFourEightN}{1}
\newcommand{\cTwoModelFableFiveTwoN}{3}
\newcommand{\cTwoMinN}{124}
\newcommand{\cTwoMinTotal}{1{,}138}
\newcommand{\cTwoMinHours}{19.0}
\newcommand{\cTwoMinMedian}{5}
\newcommand{\cTwoBktVerN}{11}
\newcommand{\cTwoBktVerSucc}{10}
\newcommand{\cTwoBktHwN}{41}
\newcommand{\cTwoBktHwSucc}{39}
\newcommand{\cTwoFoundingLogs}{8}
\newcommand{\cTwoExplBucketsSpearman}{9}
\newcommand{\cTwoExplRhoSucc}{\ensuremath{-0.48}}
\newcommand{\cTwoExplRhoSuccP}{0.22}
\newcommand{\cTwoExplSuccAtCeiling}{7}
\newcommand{\cTwoExplItMean}{2.22}
\newcommand{\cTwoExplMultiIter}{86}
\newcommand{\cTwoExplMinHwSharePct}{58.2}
\newcommand{\cTwoExplHwUnitSharePct}{31.8}
\newcommand{\cTwoExplSpanWeeks}{15}
\newcommand{\cTwoExplReviewLogs}{31}
\newcommand{\cTwoTokGenM}{57.2}
\newcommand{\cTwoTokModelsNotInLogsIds}{claude-sonnet-5-5}
\newcommand{\cTwoTokSubProcPct}{78}
\newcommand{\cTwoPredRegistered}{94}
\newcommand{\cTwoPredMet}{89}
\newcommand{\cTwoPredHashRegistered}{75}
\newcommand{\cOneN}{232}
\newcommand{\cOneSuccPct}{71.6}
\newcommand{\autoRtlLinesAuthored}{12{,}875}

\begin{document}
\bstctlcite{BSTcontrol}

\title{From FIPS to FPGA with LLM Agents:\\Measuring What Tests and Reviews Miss in Post-Quantum Cryptographic Hardware}

\author{\IEEEauthorblockN{Jungmin Park\IEEEauthorrefmark{1}, Eunha Kim\IEEEauthorrefmark{2}, Wooseop Kim\IEEEauthorrefmark{2}, Seongjoon Cho\IEEEauthorrefmark{2}, and Byungho Cha\IEEEauthorrefmark{1}}
\IEEEauthorblockA{\IEEEauthorrefmark{1}Lucid Motors, Newark, CA, USA\quad
\IEEEauthorrefmark{2}EYL Inc., Seoul, Republic of Korea}}

\maketitle

\begin{abstract}
As LLM agents increasingly generate RTL alongside the tests and reviews that validate it, self-generated evidence is a weak foundation for trustworthy hardware. We present our seven-stage verification-gated workflow, evolved during a post-quantum design port, in which a gate counts only if shown able to fail on negative controls and fault mutants, and reviewer agents audit work before commit. We then audit this evidence with tool-selected netlist mutants, flaw replays, and formal checks. The flow ported an ML-KEM-768/ML-DSA-65 accelerator to a 300\,MHz Zynq UltraScale+ key-custody card: all six operations are byte-exact on the board, and it passes all 85 interface-applicable NIST ACVP cases (of 285) and 484 ML-DSA known answers. However, passing module tests overstate the verification strength. Across three RTL modules, existing simulation gates kill only 76\% of 399 tool-selected netlist mutants; specification-derived tests guided by the survivors raise
the fresh-draw decapsulation kill rate from 48\% to 74\%. In pre-registered single-model replays on flaws invisible to known-answer tests, undirected whole-design review catches two verifier flaws in only 3/6 and 1/6 sessions, whereas module-scoped review catches every known flaw in all 54 sessions. Even scoped review detects planted
key-custody flaws in only 3/20 sessions, whereas formal checks (one added for the replay) flag every one. In this study, trustworthy agentic hardware needed author-independent mutants for test gates, module-level scoping for LLM review, and formal checks for custody invariants.
\end{abstract}


\section{Introduction}

Post-quantum migration is a hardware problem with a deadline. FIPS~203 and~204 are final~\cite{fips203,fips204}, and NIST schedules the retirement of today's public-key algorithms between 2030 and 2035~\cite{nistir8547}. Every key-holding product needs its own deployment, spanning lattice arithmetic, RTL, timing closure, a host link, host software, and board bring-up. Domain automation covers one slice, such as a kernel~\cite{perera2026llm4pqc} or an accelerator from high-level synthesis~\cite{zhang2026contracthil}, whereas large language model (LLM) agents can work across all of these artifacts with standard pseudocode and conformance vectors enabling byte-exact comparison. However, when agents generate the design, the tests and the review reports in one workflow, passing tests and approving reviews can share blind spots, and a known-answer test (KAT) cannot show that malformed input is refused or that a secret stays inside. These limitations raise a broader question: can an agentic flow port a standard-conformant PQC system onto a new FPGA target, and what makes its verification evidence trustworthy?

\begin{figure*}[t]
\centering
\begin{tikzpicture}[fg]
\newcommand{\fgStage}[6][fgbox]{\node[#1, minimum height=14.5mm, inner sep=2.6pt, anchor=north west] (#2) at (#3,0)
  {\parbox[t][12.6mm][t]{21.5mm}{\raggedright{\tiny\bfseries\color{fgSub}STAGE #4}\\[-1pt]\textbf{#5}\\[0.5pt]{\tiny\color{fgSub}#6}}};}
\fgStage{s1}{0}{1}{Oracle synthesis}{standard $\rightarrow$ executable golden model, pinned vectors}
\fgStage{s2}{2.56}{2}{Generation}{agents write or port RTL and host code; byte-exact in simulation}
\fgStage{s3}{5.12}{3}{Timing closure}{synthesis and timing reports fed back until the clock closes}
\fgStage[fgnewbox]{s4}{7.68}{4}{Gate compilation}{claim $\rightarrow$ gate + negative control + killed mutants}
\fgStage[fgnewbox]{s5}{10.24}{5}{Adversarial review}{reviewer agents and skeptics; fixer; re-gate}
\fgStage{s6}{12.80}{6}{Promotion ladder}{simulation $\rightarrow$ conformance $\rightarrow$ board, by generated runners}
\fgStage{s7}{15.36}{7}{Telemetry}{logs, tokens, operator minutes into a fixed analyzer}
\foreach \a/\b in {s1/s2,s2/s3,s3/s4,s4/s5,s5/s6,s6/s7} \draw[fgar, ->] (\a.east) -- (\b.west);
\begin{scope}[on background layer]
\node[fgzone, fit=(s2)(s5), inner sep=3.2pt] (loop) {};
\end{scope}
\node[fgeb, anchor=south west] at ([xshift=2pt,yshift=0.5pt]loop.north west) {REPEATS PER UNIT OF WORK};
\draw[fgar, ->, dash pattern=on 2.5pt off 2pt] (s6.south) -- ++(0,-0.38) -| node[pos=0.25, below, fgsub]{a failure returns to stage 2 and leaves a new gate at stage 4} (s2.south);
\end{tikzpicture}
\caption{The proposed seven-stage workflow (shaded: the two stages extended and measured here).}
\label{fig:flow}
\end{figure*}
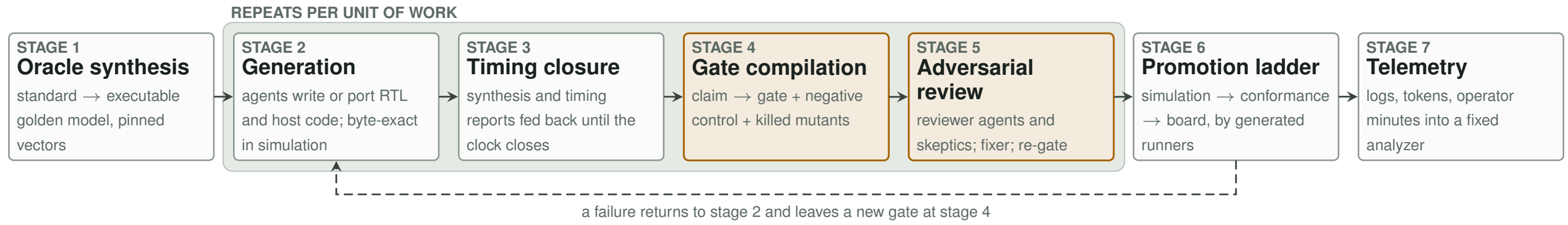

We answer with our seven-stage workflow (Fig.~\ref{fig:flow}) in which generation is cheap and evidence is the product. It fixes an executable oracle before RTL is accepted, turns each claim into a gate that can fail, has separate agents that challenge the result, and allows hardware deployment only after progressively stronger verification gates are passed.

Park et al.~\cite{park2026aiassisted} took a first target (A), an ML-KEM-768/ML-DSA-65 accelerator on a Kintex-7
card, to the board under a six-step protocol that supplies our oracle-first rule and ladder. We extend its human
review gate into adversarial review by agents and its rule per failure into gates with negative controls and
mutants. We add the generated operator runners and freeze the telemetry's analysis early.
We apply the extended flow to the second target~(B), a key-custody module on a Zynq UltraScale+ card with three times the clock and stricter requirements (Sec.~\ref{sec:target}). We claim no speed-up and no generality beyond this design family; we measure how far two
trust mechanisms hold. Our contributions are:
\begin{itemize}[leftmargin=*,nosep]
  \item \textbf{A seven-stage verification-gated workflow from standard to board conformance}, that ported an
    ML-KEM-768/ML-DSA-65 key-custody design to a 300\,MHz card that passes NIST ACVP and known-answer tests
    (Secs.~\ref{sec:flow},~\ref{sec:target}).
  \item \textbf{Failure-sensitive gates audited with tool-selected netlist mutants}: the passing gates of three modules
    kill 76\% of 399 mutants, and survivor-guided tests raise decapsulation's fresh-draw kill rate from 48\% to 74\% (Secs.~\ref{sec:gates},~\ref{sec:episodes}).
  \item \textbf{Pre-registered flaw replays of LLM review and formal custody checks}: module-scoped review
    finds every known flaw (54/54) where whole-design review often misses verifier flaws; planted custody flaws evade even scoped review (3/20) but not formal checks (one added for the
    replay) (Sec.~\ref{sec:episodes}).
  \item \textbf{A workflow evaluation with early-frozen analysis}: acceptance, iterations, minutes, tokens and
    board sessions graded against prior predictions (Secs.~\ref{sec:flow},~\ref{sec:results},~\ref{sec:threats}).
\end{itemize}

\section{Background and Related Work}

\textbf{LLMs for RTL and agentic flows.} Benchmarks and fine-tuned models target single-module
generation~\cite{liu2023verilogeval,lu2024rtllm,thakur2023benchmark,thakur2023verigen,liu2025rtlcoder,pinckney2025revisiting}, and benchmark validity
is debated~\cite{wang2025vericontaminated,ghorab2026notsotiny,wang2026rtlbenchmt}. Beyond conversational
design~\cite{blocklove2023chipchat}, agentic flows close the loop with
compilers, simulators or judge agents~\cite{thakur2023autochip,tsai2024rtlfixer,xu2024meic,ho2024verilogcoder,zhao2025mage},
or with formal equivalence to a C reference model~\cite{li2026formalrtl}, or count human
interventions~\cite{yu2025spec2rtl}. Closest in scale are an agent that took a RISC-V CPU to layout~\cite{verkor2026conductor1}, a multi-agent
harness that built an FPGA-mapped
accelerator~\cite{verkor2026conductor2} and an engineer-in-the-loop agent that designed two simple
processors~\cite{mohammed2025architect}. For post-quantum cryptography (PQC), LLMs have synthesized
kernels~\cite{perera2026llm4pqc} and co-designed a FALCON accelerator~\cite{liao2026llm4pqc}.

\textbf{Hardware In the Loop (HIL), test strength, and security review.} ContractHIL-HLS feeds board results back into a
multi-agent high-level-synthesis flow for an ML-KEM-512/ML-DSA-44 accelerator~\cite{zhang2026contracthil}; FPGAgent
runs agent-generated kernels on a board~\cite{wang2026fpgagent}. Qualifying tests by mutation is established
practice~\cite{demillo1978hints,hampton2007leveraging,mcy}; AutoBench, CorrectBench and PRO-V-R1 score LLM-written
testbenches on mutants~\cite{qiu2024autobench,qiu2025correctbench,zhao2025prov}; MARVEL and SecureRAG-RTL apply LLM agents to security
review of RTL~\cite{collini2025marvel,hasan2026securerag}; HWE-Bench traces agent failures on real hardware bugs partly to
fault localization~\cite{cui2026hwebench}; two flows use known-answer tests (KATs) in
simulation~\cite{perera2026llm4pqc,verkor2026conductor2}. We combine these ideas into one
flow with acceptance rules, applied at system scale and graded on hardware against standard conformance vectors
(Table~\ref{tab:related}).

\begin{table}[t]
\caption{Closest related flows.}
\label{tab:related}
\centering\scriptsize
\setlength{\tabcolsep}{2.5pt}
\begin{tabularx}{\columnwidth}{@{}>{\raggedright\arraybackslash}Xlllll@{}}
\toprule
Work & Level & HW & Std.\ vectors & Sec. & Cost \\
\midrule
ContractHIL-HLS~\cite{zhang2026contracthil} & HLS, PQC & board & -- & -- & tokens \\
FPGAgent~\cite{wang2026fpgagent} & HLS kernels & board & -- & -- & -- \\
LLM4PQC~\cite{perera2026llm4pqc} & HLS, kernels & -- & KAT, sim. & -- & -- \\
Design Conductor~2~\cite{verkor2026conductor2} & RTL & mapped & KAT, sim.$^{a}$ & -- & rel.\ tokens \\
MARVEL~\cite{collini2025marvel} & RTL review & -- & -- & yes & \$/run \\
Park et al.~\cite{park2026aiassisted} & RTL, PQC & board & KAT, board & -- & tokens \\
This flow & RTL port, PQC & board & ACVP$^{b}$, board & yes & tokens, min. \\
\bottomrule
\end{tabularx}\\[2pt]
\parbox{\columnwidth}{\scriptsize From each paper's full text; ``--'': not reported. HW: run on a board, or only mapped. Std.\ vectors: graded against a
standard's published vectors. Sec.: a security review stage. Cost: what is reported. $^{a}$AES core only. $^{b}$85 of 285 cases.}
\end{table}

\textbf{The base protocol.} Park et al.~\cite{park2026aiassisted} took the first target from RTL to a working
PCIe card under six steps: (P1) an executable golden reference first, byte-exact agreement being the sole
acceptance criterion; (P2) a versioned constitution file; (P3) a structured log per unit of work; (P4) a human review gate on every artifact;
(P5) a typed memory rule for every root-caused failure; (P6) promotion
only up a four-rung ladder. They publish it as a repository template in which only the constitution and the
reference directory change per target, report \cOneSuccPct\% task acceptance over
\cOneN{} logged experiments (counted differently from our rate; Sec.~\ref{sec:threats}), and state that transfer to another target was untested.

\section{The Proposed Workflow}
\label{sec:flow}

Our workflow (Fig.~\ref{fig:flow}) takes a standard, requirements and the target board's documentation,
and produces a bitstream, the host software that drives it and a manifest of evidence.

Work is cut into \emph{units}. A unit is one deliverable: a hardware module, host-software function, test
harness, build, board session, document or plan. A unit passes through those of stages~2--5 that apply to it and is logged once at stage~7.
This section describes the flow as it stood at the end: stages~4 and~5 took shape during the port.
Of the \cTwoN{} logs, \cTwoExplReviewLogs{} record a review verdict, none from the first month; the logs do not
record which units passed a gate with mutants.

\textbf{Stage 1: oracle synthesis.} Design starts from a software model of each scheme, the \emph{oracle}: here public
reference implementations and the project's own models. An output is correct only if it equals the oracle's
output byte for byte. The oracle is itself checked against the vectors published with the standard; on B this came
late (285 NIST ACVP cases, all reproduced). Where the device cannot take a published vector as is, the stage makes known answers in the device's own
format, on which two independent implementations agree.

\textbf{Stage 2: generation of hardware and host software.} Builder agents write or port one piece per unit. For
hardware, the testbench compares each module with the oracle's intermediate values. The host side (libraries, a
custody daemon, test tools) is built the same way. Host code first runs against a software model of the device, faults
included.

\textbf{Stage 3: timing closure.} Agents read the synthesis and timing reports and rewrite the failing path, here by
registering the ports of the shared memory and the operation select, and later by rewriting constant multiplications
as shift-and-add. Build checks catch what timing analysis cannot: every
constant ROM is initialized and the PCIe lane order matches the board.

\textbf{Stage 4: gate compilation.} \label{sec:gates}Every claim of a unit, such as ``the verifier rejects a signature with a malformed hint field'', becomes a \emph{gate}, an automatic check. When
one kind of agent writes the design and its test, a test can pass because it never looks. So each gate gets a
\emph{negative control}, an input on which it must fail, which exposes a check that never fires, and \emph{mutants},
small breaks of the code it guards, which expose a check that fires only for the case its author thought
of. A gate is accepted only if it passes on the real code and fails on every mutant. In their last runs
the host libraries' gates killed 402 of 402 mutants and the regression runner's 53 of 53: 100\% by rule, as a
gate is reworked until its agent-written mutants die.
Sec.~\ref{sec:episodes} measures what gates miss, on faults no agent chose.

\textbf{Stage 5: adversarial review.} Before a unit is committed, reviewer agents check it again, each from one named
viewpoint (a \emph{lens}, such as correctness, security, device safety, or claims and records), and return a verdict with findings.
Skeptic agents then try to disprove each finding. A fixer agent repairs the confirmed ones, and the gates run again.

\textbf{Stage 6: promotion ladder.} A unit climbs a ladder: byte-exact in simulation, the conformance
vectors, byte-exact on the board, then a soak run (on B: one hour, three operations, an earlier build). For each board session the flow
writes a \emph{runner}, a one-command script for the operator. It refuses an unsafe state, drives the host tools and writes an \emph{evidence manifest}, a list of the session's files with a hash of each, so that none can change unnoticed. For later sessions the agent wrote numeric predictions first; grades come from
these files, not the terminal.

\textbf{Stage 7: telemetry.} Every unit closes with a structured log; a tool records token use at every commit. To keep the analysis from being fitted to the results, we froze it early~\cite{nosek2018preregistration,baltes2026guidelines}: the
categories, buckets, bucket order and analysis script were committed nine days into the work, before the core was
ported. The script that produces our numbers checks that they are unchanged.
The \cTwoFoundingLogs{} units logged by then, three of them technical, are among the \cTwoN. A unit counts as accepted even after several iterations, so acceptance is not a first-attempt rate.

\textbf{Context and prompts.}\label{sec:context} What an agent must know is written in files: the constitution
(acceptance criterion, commands, prohibitions; rewritten per target), golden models and pinned vectors, one record
per work package and one memory rule per root-caused failure. Sub-agent prompts follow a common pattern: the record
to read first, binding rules, the files the agent may edit, the task as properties the result must have, the
validation to run, and an output form, usually fixed JSON. A reviewer never edits, gets one lens and treats the
builder's report as claims to check.

\section{Two Targets and Two Requirement Sets}
\label{sec:target}

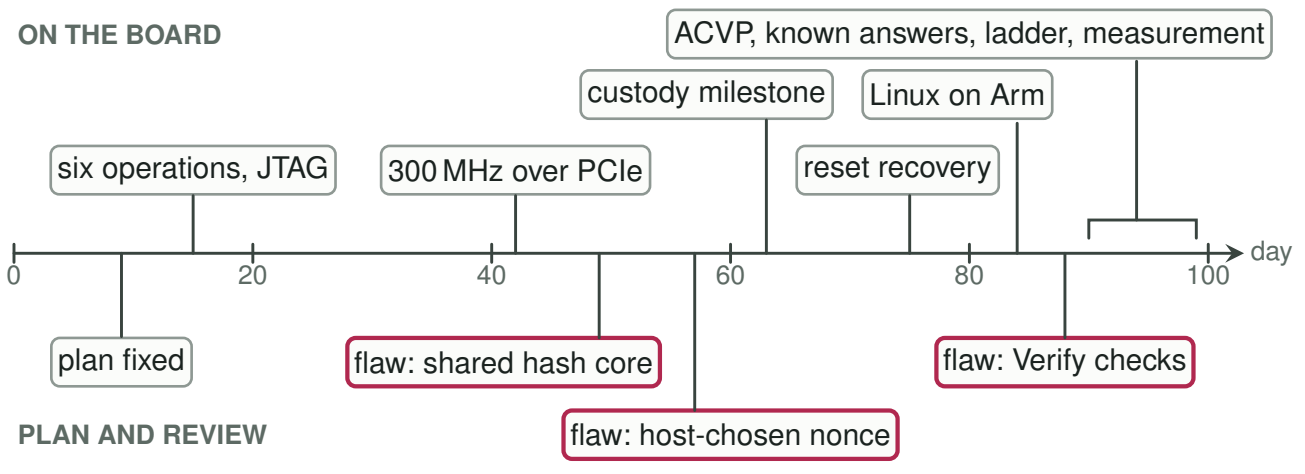
\begin{figure}[t]
\centering
\begin{tikzpicture}[fg, x=0.079cm, y=1cm,
  ev/.style={fgbox, font=\sffamily\fontsize{5.6}{6.4}\selectfont, inner sep=1.3pt, minimum height=3.2mm},
  fl/.style={ev, draw=fgKey, line width=0.8pt}]
\draw[fgar, ->] (0,0) -- (103,0) node[right, fgsub, inner sep=1pt] {day};
\foreach \d in {0,20,40,60,80,100} { \draw[fgar] (\d,-0.06) -- (\d,0.06); \node[fgsub, below=1.5pt, inner sep=0.5pt] at (\d,0) {\d}; }
\node[fgeb, anchor=west] at (0,1.45) {ON THE BOARD};
\node[ev] (a) at (15,0.55) {six operations, JTAG};
\node[ev] (b) at (42,0.55) {300\,MHz over PCIe};
\node[ev] (c) at (58,1.05) {custody milestone};
\node[ev] (d) at (74,0.55) {reset recovery};
\node[ev] (e) at (79,1.05) {Linux on Arm};
\node[ev] (f) at (80,1.45) {ACVP, known answers, ladder, measurement};
\draw[fgar] (15,0) -- (a.south); \draw[fgar] (42,0) -- (b.south);
\draw[fgar] (63,0) -- (63,0.86) -- (c.south -| 63,0.86);
\draw[fgar] (75,0) -- (75,0.36) -- (d.south -| 75,0.36);
\draw[fgar] (84,0) -- (84,0.86);
\draw[fgar] (90,0.1) -- (90,0.22) -- (99,0.22) -- (99,0.1); \draw[fgar] (94,0.22) -- (94,1.27);
\node[fgeb, anchor=west] at (0,-1.2) {PLAN AND REVIEW};
\node[ev] (p) at (9,-0.72) {plan fixed};
\node[fl] (g) at (41,-0.72) {flaw: shared hash core};
\node[fl] (h) at (60,-1.2) {flaw: host-chosen nonce};
\node[fl] (i) at (88,-0.72) {flaw: Verify checks};
\draw[fgar] (9,0) -- (p.north); \draw[fgar] (49,0) -- (49,-0.56);
\draw[fgar] (57,0) -- (57,-1.04); \draw[fgar] (88,0) -- (88,-0.56);
\end{tikzpicture}
\caption{Logged work on target~B, in days from its first unit.}
\label{fig:timeline}
\end{figure}

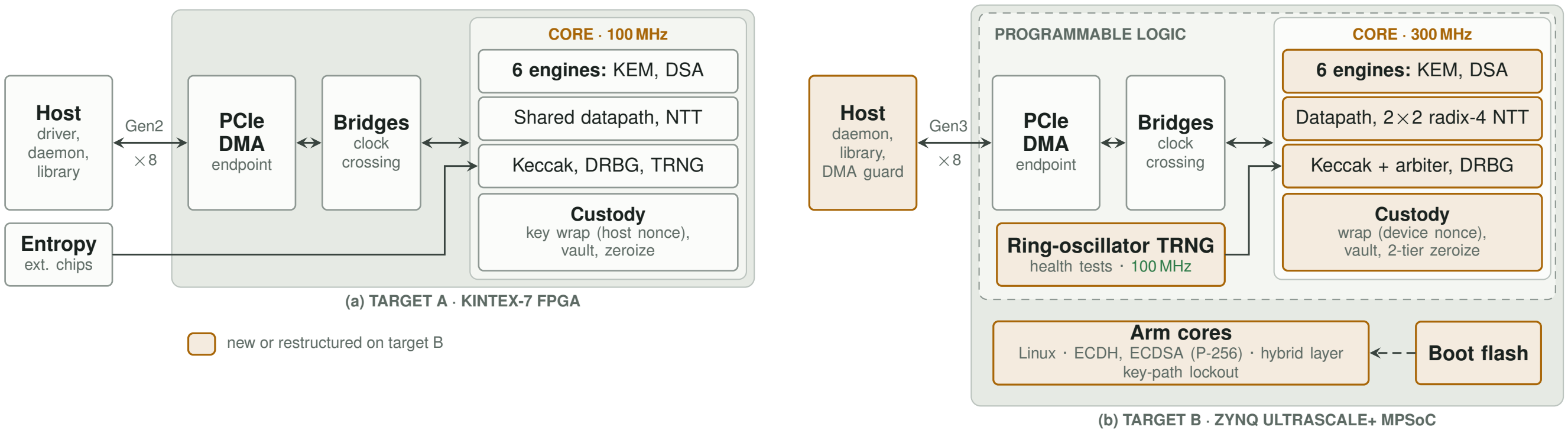
\begin{figure*}[t]
\centering
\begin{tikzpicture}[fg,
  core/.style={fgbox, font=\sffamily\fontsize{6}{7}\selectfont, text width=28mm, minimum width=29mm, inner sep=1.2pt},
  corenew/.style={core, draw=fgNew, fill=fgNew!13, line width=0.6pt},
  side/.style={fgbox, font=\sffamily\fontsize{6.5}{7.5}\selectfont, inner sep=1.2pt}]
\node[side, text width=11mm, minimum width=12mm, minimum height=15mm] (ah) at (0.70,0) {\fgT{Host}{driver,}\\[-1.5pt]\fgS{daemon,}\\[-1.5pt]\fgS{library}};
\node[side, text width=11mm, minimum width=12mm, minimum height=7mm] (aq) at (0.70,-1.25) {\fgT{Entropy}{ext. chips}};
\node[side, text width=11mm, minimum width=12mm, minimum height=15mm] (ad) at (2.75,0) {\fgT{PCIe DMA}{endpoint}};
\node[side, text width=10mm, minimum width=11mm, minimum height=15mm] (aw) at (4.20,0) {\fgT{Bridges}{clock}\\[-1.5pt]\fgS{crossing}};
\node[core] (ae) at (6.85,0.80) {\textbf{6 engines:} KEM, DSA};
\node[core] (an) at (6.85,0.27) {Shared datapath, NTT};
\node[core] (ak) at (6.85,-0.26) {Keccak, DRBG, TRNG};
\node[core, minimum height=8.6mm] (ac) at (6.85,-1.0) {\fgT{Custody}{key wrap (host nonce),}\\[-1.5pt]\fgS{vault, zeroize}};
\node[fgeb, text=fgNew] (acl) at (6.85,1.22) {CORE \textperiodcentered{} 100\,MHz};
\begin{scope}[on background layer]
\node[fgzone, fit=(ad)(aw)(ae)(ac)(acl), inner sep=5pt] (afp) {};
\node[fginner, fit=(ae)(an)(ak)(ac)(acl), inner sep=2.5pt] (acore) {};
\end{scope}
\node[fgeb, below=1.5pt of afp] {(a) TARGET A \textperiodcentered{} KINTEX-7 FPGA};
\draw[fgar, <->] (ah) -- node[above, fgsub, pos=0.42]{Gen2} node[below, fgsub, pos=0.42]{$\times$8} (ad);
\draw[fgar, <->] (ad) -- (aw);
\draw[fgar, <->] (aw) -- (acore.west |- aw);
\draw[fgar, ->] (aq.east) -- (5.02,-1.25) |- (ak.west);
\node[fgnewbox, minimum width=3mm, minimum height=2.4mm, inner sep=0pt] (lg) at (2.3,-2.25) {};
\node[fgsub, anchor=west] (lgt) at (lg.east) {new or restructured on target B};
\begin{scope}[xshift=9.0cm]
\node[side, draw=fgNew, fill=fgNew!13, line width=0.6pt, text width=11mm, minimum width=12mm, minimum height=15mm] (bh) at (0.70,0) {\fgT{Host}{daemon,}\\[-1.5pt]\fgS{library,}\\[-1.5pt]\fgS{DMA guard}};
\node[side, text width=11mm, minimum width=12mm, minimum height=15mm] (bd) at (2.75,0) {\fgT{PCIe DMA}{endpoint}};
\node[side, text width=10mm, minimum width=11mm, minimum height=15mm] (bw) at (4.20,0) {\fgT{Bridges}{clock}\\[-1.5pt]\fgS{crossing}};
\node[corenew] (be) at (6.85,0.80) {\textbf{6 engines:} KEM, DSA};
\node[corenew] (bn) at (6.85,0.27) {Datapath, 2$\times$2 radix-4 NTT};
\node[corenew] (bk) at (6.85,-0.26) {Keccak + arbiter, DRBG};
\node[corenew, minimum height=8.6mm] (bc) at (6.85,-1.0) {\fgT{Custody}{wrap (device nonce),}\\[-1.5pt]\fgS{vault, 2-tier zeroize}};
\node[fgeb, text=fgNew] (bcl) at (6.85,1.22) {CORE \textperiodcentered{} 300\,MHz};
\node[side, draw=fgNew, fill=fgNew!13, line width=0.6pt, text width=24.5mm, minimum width=25.5mm, minimum height=7mm] (bt) at (3.475,-1.25) {\fgT{Ring-oscillator TRNG}{health tests \textperiodcentered{} {\color{fgGreen}100\,MHz}}};
\node[fgeb, anchor=west] (bpll) at (2.15,1.22) {PROGRAMMABLE LOGIC};
\node[side, draw=fgNew, fill=fgNew!13, line width=0.6pt, text width=41mm, minimum width=42mm, minimum height=7mm, anchor=west] (bps) at (2.15,-2.35) {\fgT{Arm cores}{Linux \textperiodcentered{} ECDH, ECDSA (P-256) \textperiodcentered{} hybrid layer}\\[-1.5pt]\fgS{key-path lockout}};
\node[side, draw=fgNew, fill=fgNew!13, line width=0.6pt, text width=13mm, minimum width=14mm, minimum height=7mm, anchor=east] (bfl) at (8.30,-2.35) {\textbf{Boot flash}};
\begin{scope}[on background layer]
\node[fgzone, fit=(bd)(bw)(be)(bc)(bcl)(bt)(bpll)(bps)(bfl), inner sep=6.5pt] (bsoc) {};
\node[fginner, draw=fgBoxLn, dash pattern=on 2.5pt off 2pt, fit=(bd)(bw)(be)(bc)(bcl)(bt)(bpll), inner sep=4pt] (bpl) {};
\node[fginner, fill=fgBox, fit=(be)(bn)(bk)(bc)(bcl), inner sep=2.5pt] (bcore) {};
\end{scope}
\node[fgeb, below=1.5pt of bsoc] {(b) TARGET B \textperiodcentered{} ZYNQ ULTRASCALE+ MPSoC};
\draw[fgar, <->] (bh) -- node[above, fgsub, pos=0.42]{Gen3} node[below, fgsub, pos=0.42]{$\times$8} (bd);
\draw[fgar, <->] (bd) -- (bw);
\draw[fgar, <->] (bw) -- (bcore.west |- bw);
\draw[fgar, ->] (bt.east) -- (5.02,-1.25) |- (bk.west);
\draw[fgar, ->, dash pattern=on 2.5pt off 2pt] (bfl.west) -- (bps.east);
\end{scope}
\end{tikzpicture}
\caption{Architecture of the two targets.}
\label{fig:arch}
\end{figure*}

\textbf{Requirements.} A is a Kintex-7 160T card with SPI flash~\cite{park2026aiassisted}; B is a Zynq UltraScale+
ZU7EV card with four Cortex-A53 cores, 4\,GB of DDR4 and QSPI flash. B's whole design, PCIe DMA endpoint included,
takes 94,505 LUTs (41.0\%), with 0.151\,ns of setup slack
on the 300\,MHz core clock; A took 88,478 LUTs (87.3\%). Table~\ref{tab:req} gives the requirements and the gate showing
each on~B. B, our measured target, was ported
from A's RTL; some of its modules were prototyped before the first logged unit. Figs.~\ref{fig:timeline}
and~\ref{fig:arch} give the timeline and both designs (panel~(b) made by the flow).

\begin{table}[t]
\caption{Functional requirements of the two targets.}
\label{tab:req}
\centering\scriptsize
\setlength{\tabcolsep}{2.5pt}
\begin{tabularx}{\columnwidth}{@{}>{\raggedright\arraybackslash}p{1.3cm}>{\raggedright\arraybackslash}p{1.25cm}>{\raggedright\arraybackslash}X>{\raggedright\arraybackslash}p{2.2cm}@{}}
\toprule
Asked for & Target A & Target B & Gate on B \\
\midrule
Core clock & 100\,MHz & 300\,MHz & timing; 6/6 byte-exact \\
Host link & PCIe Gen2\,$\times$8 & PCIe Gen3\,$\times$8 & lane-order gate; link on the board \\
Conform. & known answers & NIST ACVP and known answers, on the device & 85/85 (of 285); 484/484 \\
Zeroization & on command & two tiers (operational, full) & 49/49 \\
Entropy & external + on-chip & on-chip; health tests fail-closed & 7/7 on the generator output; fail-closed in simulation \\
Host reset & function reset & bus reset, no power cycle & 13/13$^{a}$ \\
Software & host library & plus Linux, ECC/RSA and a hybrid layer on the Arm cores & 119/119 (device stubbed); 6/6$^{a}$; RSA not built \\
Boot & from flash & from flash via the processors, locked out of the key path & lock-out 5/5 over JTAG; flash open$^{b}$ \\
\bottomrule
\end{tabularx}\\[2pt]
\parbox{\columnwidth}{\scriptsize $n/m$: $n$ checks passed of $m$. $^{a}$On the board, on an earlier
build of B. $^{b}$Never booted from flash: the backup before the write failed (the vendor tool cannot read this flash back).
Over JTAG the lock-out boot chain's fourth session passed (three stopped early): Linux, six operations 6/6, all five lock-out probes refused (bench image, one run).}
\end{table}

\textbf{Evidence on the board.} \emph{Integration}: the custody daemon (49 of 49 checks) and, with P-256 ECDH and ECDSA in software, the hybrid KEM and dual signature (15 of 15 from the host; 119 of 119 on the Arm cores, device stubbed), all on an earlier build.
A fail-closed guard on the DMA output passed 31 of 31.
\emph{Verification}: all 85 ACVP cases the device interface can express (of 285; KeyGen cases compare the public key
only; 20 more are host-library key checks). The device signs and verifies only 32-byte messages with an empty
context, deterministically, without pre-hash or external $\mu$; none of the 180 ACVP signing and verification cases
has that form. Instead: 484 ML-DSA-65 known answers on which two independent implementations agree, 121 signatures and 363
verifications (121 valid, 121 with a modified message, 121 under a wrong key). A 12-step regression ladder was
graded ``pass with carve-outs'' (one custody check fails on Sec.~\ref{sec:episodes}'s wrapping-key limit; one run).
\emph{Measurement}: on the final build (one session; 26{,}346 byte-exact operations over all paths) the C
library's fast path (PCIe, one stream) runs ML-KEM-768 KeyGen, Encaps and Decaps at 10,391, 7,100 and 4,129
operations/s, and ML-DSA-65 KeyGen, Sign and Verify at 3,339, 2,565 and 4,035 (Sign in one attempt; 1,031 on
average, 5.38 attempts). That is 3.0--5.1$\times$ A's fast path~\cite{park2026aiassisted} (Sign
5.3$\times$; few or no rejections on~A). The default path resets before each
operation, like the conformance runs, and drains for 3\,ms: 281--305 operations/s (average signing 242).

\section{Evaluation of Our Workflow}
\label{sec:results}

Unit counts and acceptance come from the analysis script frozen at stage~7; the other figures are exploratory.

\textbf{Acceptance and effort.} Of \cTwoN{} units, \cTwoSucc{} were accepted (\cTwoSuccPct\%, Wilson 95\% CI
\cTwoWilsonLo--\cTwoWilsonHi\%). The plan sorts units into task buckets from least to most hardware-coupled; the
three failures are in verification (\cTwoBktVerSucc{} of \cTwoBktVerN{} accepted) and board bring-up (\cTwoBktHwSucc{} of
\cTwoBktHwN), the largest bucket. \cTwoOneShot{}
(\cTwoOneShotPct\%) closed after one iteration and \cTwoExplMultiIter{} took two or more (mean
\cTwoExplItMean, maximum \cTwoIterMax); \cTwoEditFree{} (\cTwoEditFreePct\%) record no human edit. A machine oracle graded \cTwoOracleMachine{} units (\cTwoOracleMachinePct\%),
\cTwoOracleHw{} on the board. The plan's one hypothesis, that acceptance falls with hardware coupling~\cite{park2026aiassisted},
could not be tested: every unit was accepted in \cTwoExplSuccAtCeiling{} of the \cTwoExplBucketsSpearman{} buckets with units, so the rates hardly differ
(Spearman $\rho=\cTwoExplRhoSucc$, $p=\cTwoExplRhoSuccP$). Coupling tracks operator time instead (exploratory): bring-up holds
\cTwoExplMinHwSharePct\% of operator minutes on \cTwoExplHwUnitSharePct\% of units.

\begin{figure}[t]
\centering
\begin{tikzpicture}[fg,
  cb/.style={fgbox, font=\sffamily\fontsize{6.3}{7.3}\selectfont, inner sep=1.2pt, minimum height=7mm},
  ck/.style={cb, draw=fgKey, line width=0.9pt}]
\node[fgeb, anchor=west] (e1) at (2.44,4.74) {1 \textperiodcentered{} PROVISION};
\node[fgeb, anchor=east] (ed) at (8.44,4.74) {DEVICE};
\node[ck, text width=13.6mm, minimum width=14.4mm] (seed) at (3.14,4.15) {\fgT{Seed}{DRBG or host}};
\node[ck, text width=13.6mm, minimum width=14.4mm] (kg)   at (4.70,4.15) {\fgT{KeyGen}{on the device}};
\node[ck, text width=27mm, minimum width=28mm] (wrap) at (7.03,4.15) {\fgT{Key wrap}{KEK(device ID) \textperiodcentered{} device nonce}};
\node[ck, text width=22mm, minimum width=23mm] (hash) at (7.31,3.10) {\fgT{Hash core + arbiter}{DRBG, engines, (un)wrap}};
\node[fgeb, anchor=west] (e2) at (2.42,2.62) {2 \textperiodcentered{} USE: SIGN OR DECAPS FROM THE VAULT};
\node[cb, text width=13.6mm, minimum width=14.4mm] (vault) at (3.14,1.60) {\fgT{Vault RAM}{blob staged}};
\node[ck, text width=13.6mm, minimum width=14.4mm] (unw)   at (4.68,1.60) {\fgT{Unwrap}{KEK, tag check}};
\node[ck, text width=13.6mm, minimum width=14.4mm] (sk)    at (6.22,1.60) {\fgT{Key RAM}{plaintext key}};
\node[ck, text width=13.6mm, minimum width=14.4mm] (use)   at (7.76,1.60) {\fgT{Sign/Decaps}{reads key RAM}};
\node[cb, text width=36mm, minimum width=37mm] (zero) at (6.58,0.30) {\fgT{Zeroize after sign, decaps, wrap}{key RAM, keystream, work memory}};
\node[fgeb, anchor=west] (eh) at (0.12,4.74) {HOST};
\node[cb, text width=16.6mm, minimum width=17.4mm] (hp) at (1.0,4.15) {\fgT{Provision}{ML-KEM: host seed, z}};
\node[cb, text width=16.6mm, minimum width=17.4mm] (hr) at (1.0,3.10) {\fgT{Record}{blob + public key}};
\node[cb, text width=16.6mm, minimum width=17.4mm, minimum height=13mm] (hc) at (1.0,1.60) {\fgT{Client}{blob to the vault;}\\[-1.5pt]\fgS{digest or ciphertext}\\[-1.5pt]\fgS{to the engine}};
\node[ck, minimum width=2.4mm, minimum height=2.4mm, inner sep=0pt] (lg) at (0.32,0.30) {};
\node[fgsub, anchor=west, inner sep=2pt] (lgt) at (lg.east) {secret material};
\begin{scope}[on background layer]
\node[fgzone, fit=(eh)(hp)(hc)(lg)(lgt), inner sep=3.5pt] (hz) {};
\node[fgzone, fit=(e1)(ed)(wrap)(use)(zero)(vault), inner sep=3.5pt] (dz) {};
\end{scope}
\draw[fgbound] ($(hz.north east)!0.5!(dz.north west)+(0,0.06)$) -- ($(hz.south east)!0.5!(dz.south west)+(0,-0.06)$);
\node[fgsub, text=fgKey, anchor=north] at ($(dz.south)+(0,-0.04)$) {library flows: no plaintext private key crosses this boundary};
\draw[fgar, ->] (hp) -- (seed);
\draw[fgar, ->] (seed) -- (kg);
\draw[fgar, ->] (kg) -- (wrap);
\draw[fgar, <->, densely dotted] (hash.north) -- (hash.north |- wrap.south);
\draw[fgar, ->] ([xshift=-12mm]wrap.south) |- node[pos=0.75, above, fgsub]{blob + public key} (hr.east);
\draw[fgar, ->] (hc.east |- vault) -- (vault);
\draw[fgar, ->] (vault) -- (unw);
\draw[fgar, ->] (unw) -- (sk);
\draw[fgar, ->] (sk) -- (use);
\draw[fgar, ->] ([xshift=-2mm]use.north) -- ++(0,0.17) -- node[pos=0.42, above, fgsub]{signature or shared secret} (hc.east |- 0,2.12);
\draw[fgar, ->] ([xshift=3mm]use.south) -- ([xshift=3mm]use.south |- zero.north);
\draw[fgar, ->, dash pattern=on 2.5pt off 2pt] (zero.north -| sk) -- node[left, fgsub, pos=0.45]{wipes} (sk.south);
\node[fgmk] at ([xshift=-1.4mm,yshift=-0.3mm]hash.north east) {1};
\node[fgmk] at ([xshift=-1.4mm,yshift=-0.3mm]wrap.north east) {2};
\end{tikzpicture}
\caption{Key custody on target~B (circled 1, 2: sites of A's custody flaws).}
\label{fig:custody}
\end{figure}
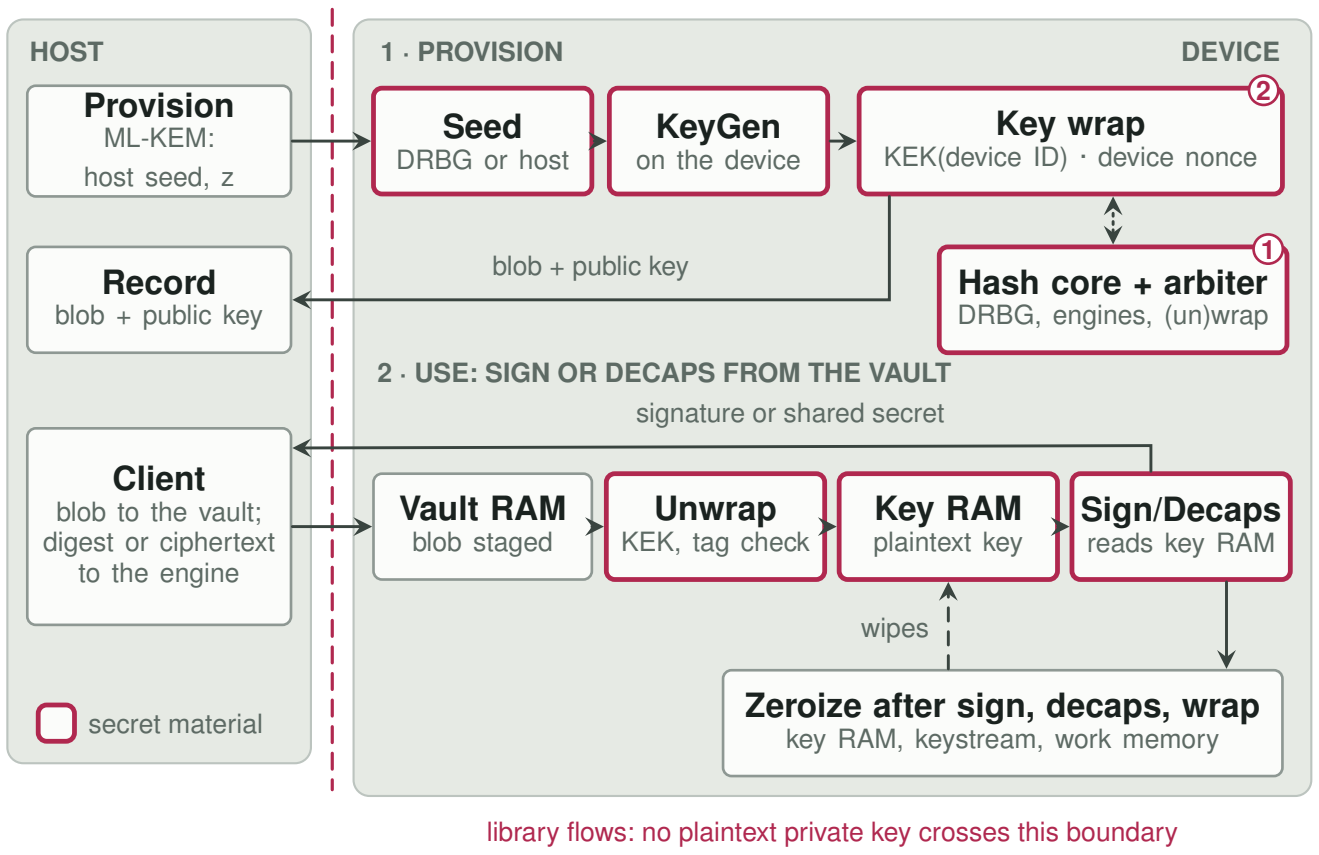

\begin{table*}[t]
\caption{Selected episodes: what got past the generating agents.}
\label{tab:episodes}
\centering\scriptsize
\setlength{\tabcolsep}{3pt}
\begin{tabularx}{\textwidth}{@{}l>{\raggedright\arraybackslash}X>{\raggedright\arraybackslash}p{4.1cm}l>{\raggedright\arraybackslash}p{4.3cm}@{}}
\toprule
 & What got past the agent & Caught by & Delay & Lasting change \\
\midrule
E1 & A constant ROM was zeroed (init file not a build source); every build closed timing & first run over PCIe: 4 of 6 byte-exact & ${\sim}$3.5 weeks & post-synthesis gate on ROM loads \\
E2$^{\dagger}$ & Verify lacked reject checks that only malformed signatures reach; all known answers passed & conformance review asking which checks the ACVP vectors reach & since before the port & boundary testbenches and mutants on the added checks \\
E2b$^{\dagger}$ & Two custody flaws from target~A let a hostile host recover a wrapped key: generator and key wrap shared a hash core without exclusion; the host chose the wrap nonce & an audit agent with simulation probes (hash core); a redesign, then security reviewers (nonce) & since before the port & grant-exclusion invariant gate; device-chosen nonce \\
E3 & A bitstream that met timing on every clock but returned no DMA output & promotion ladder on the board (vs.\ the previous bitstream) & 1 day & DMA smoke test plus full ladder before promotion \\
E4$^{\dagger}$ & A hybrid KEM combiner omitted a required binding; its gates were blind to two regressions & review rounds 1--2, before the board & same day & mutation gates that must fail first \\
E5 & A host-library Verify fell back to a 30-s poll; all 484 correct & pre-committed time prediction (missed) & ${\sim}$7 weeks & completion on the done latch; re-run 75\,s \\
E6$^{\dagger}$ & Boot firmware that would have re-configured four processor pins, accepted after three review rounds & reviewer agents in a later unit (never booted) & 14 days & boot image refused without the gate record \\
E7$^{\dagger}$ & A flash-update runner whose recovery could not run after a failed boot and could save the device's own image as the factory one; its suite killed all 66 mutants & two review agents independently, confirmed by a skeptic (before any board run) & same day & write-once factory record; recovery cases in its suite \\
\bottomrule
\end{tabularx}\\[2pt]
\parbox{\textwidth}{\scriptsize Delay: from the error entering an artifact or record to its detection. $^{\dagger}$Found by LLM agents alone (audit, review or redesign), with every test passing.}
\end{table*}

\textbf{Cost.}\label{sec:cost} Each log records the operator's minutes on its unit (specifying, answering, reviewing, acting at the board),
estimated by the agent at close-out. Waiting while agents worked and work between units
are not counted, and the operator recalls long waits, so the figures are a lower bound. \cTwoMinN{} of the \cTwoN{} logs record \cTwoMinTotal{} minutes in all (\cTwoMinHours\,h over \cTwoExplSpanWeeks{} weeks;
median \cTwoMinMedian{} per unit). The port changed or wrote \autoRtlLinesAuthored{} RTL lines, about 1,600 of them
prototyped before the first logged unit; with no human-only run we claim no speed-up. The token ledger covers the
repository's sessions after the first days: \cTwoTokGenM\,M generated tokens (about 15\% on days with no unit of B;
one session holding 15 units is missing), at least \cTwoTokSubProcPct\% of processed tokens from parallel sub-agents.

\section{Measuring Gates and Review}
\label{sec:episodes}

Table~\ref{tab:episodes} samples eight episodes, chosen by us and not a census: in each, something wrong got past the
generating agents, a later stage caught it, and the fix left a lasting gate or rule. In the five rows marked
$\dagger$, every test passed and LLM agents found the flaws.
The rest measures two extended stages: gates (stage~4) on two tool-mutant draws, review (stage~5) in two replays.
All four protocols were \emph{pre-registered}: fixed and hashed on our own machine before their data were drawn, with
no third-party timestamp. Deviations are archived; the first replay was re-run in full after a model change stopped
it.

\begin{table}[t]
\caption{Percentage of mutants killed by the gates (higher is better).}
\label{tab:gates}
\centering\scriptsize
\setlength{\tabcolsep}{2.5pt}
\begin{tabularx}{\columnwidth}{@{}>{\raggedright\arraybackslash}Xlrrrr@{}}
\toprule
Mutants & Gates & Transform & Verifier & Decaps. & All \\
\midrule
399 tool mutants (first draw) & measured & 92\% & 67\% & 53\% & 76\% \\
150 tool mutants (fresh draw) & measured & 88\% & 76\% & 48\% & 70.7\% \\
the same 150 & measured + new & 88\% & 82\%$^{a}$ & 74\%$^{b}$ & 81.3\%$^{c}$ \\
\bottomrule
\end{tabularx}\\[2pt]
\parbox{\columnwidth}{\scriptsize Example: the measured gates failed on 67 of the verifier's 100 first-draw
mutants. Modules: B's radix-4 transform, ML-DSA verifier and ML-KEM decapsulation, with 199, 100 and 100 first-draw
mutants (1 of 400 not writable as Verilog) and 50 each in the fresh draw; All pools them, so the first draw weights
the transform double (equal weights: 70.7\%). A time-out counts as a kill.
An audit confirmed every kill count and corrected the survivors' equivalence check. Fresh draw: measured-gate rates
agree with the first draw (Fisher $p\ge0.34$); 2 of its 150 mutants repeat first-draw ones, neither a gain. New tests
use ACVP vectors, FIPS~203/204 models and module-header
properties; each passes on the original and fails on its negative control. Gain, the mutants killed only with the new
tests (none can be lost): $^{a}$3 of 50 (Wilson 95\% CI 2--16\%); $^{b}$13 of 50 (16--40\%); $^{c}$16 of 150
(7--17\%). At module ports only (post hoc): $^{a}$1 and $^{b}$10 of 50; the other gains are internal-register checks
and one compile failure. Agent-written mutants (30 per module, exploratory, writer not blind to the protocol): 93\%,
87\% and 70\% killed; none differs significantly from the first draw (Holm). Of this experiment's five predictions,
four held (one only by the protocol's rule, not at module ports).}
\end{table}

\textbf{How much the gates miss in stage 4.} Agents write stage~4's mutants, so killing them all shows that each gate
can fail, not how many faults the gates miss. So we followed established mutation testing~\cite{demillo1978hints} with faults no agent chose. A
\emph{tool mutant} is a module with one random bit of its netlist changed by Yosys
\texttt{mutate}, the command MCY is built on~\cite{mcy}. It is \emph{killed} when a gate fails on it. A survivor passes
all measured gates: they miss that
fault, unless it behaves like the original and cannot be killed. We measured three modules on two or three simulation gates each, chosen before any mutant was drawn. We left out the two
custody modules: their whole-design gates did not fit a budget of about 90 seconds per mutant. Only one measured gate has
stage-4 mutants of its own on record; no other simulation or board gate was run. The measured gates killed 303 of 399 tool mutants
(76\%; \emph{first draw}, Table~\ref{tab:gates}). Of the 96 survivors, 19 provably equal the original, 22
provably differ and 55 are undecided, so the gates killed 80--93\% of the mutants that
change behaviour. Survivors located real gaps. The verifier's hint gate, run last,
killed none of the 33 mutants that reached it. No measured decapsulation gate exercised ML-KEM's \emph{implicit rejection}, its
pseudo-random answer to a bad ciphertext; one system-level zeroization test does. Single-bit mutants do not model
missing logic, a class among the known flaws below, so they cannot show whether tests catch it.

\textbf{New tests for the gaps in stage 4.} Two of five new tests raised decapsulation's kill rate
from 48\% to 74\% (Table~\ref{tab:gates}). Guided by first-draw survivors, one agent wrote all five in 1\,h 43\,min,
with expected values from the specification, not RTL simulation. All pass on the unchanged design: a surviving mutant marks a gap in the
tests, not an RTL bug. They were frozen before the \emph{fresh draw} of 50 tool
mutants per module; the writer knew the draw's seed (Sec.~\ref{sec:threats}). Decapsulation's gain,
mostly on implicit rejection, is large; the verifier's is small and the transform's nil. The known flaws below were found by
agent review and audit (Table~\ref{tab:episodes}), not by mutants.

\textbf{Flaws that known-answer tests cannot show.} A known-answer test compares outputs for well-formed input,
so it cannot show that malformed input is refused or that a secret stays inside. E2 and E2b hold 4 such flaws, the
\emph{known flaws}, inherited from target~A. The adversary is host software with full access to the device's registers
and data streams. Physical access, side channels, fault injection, JTAG and bitstream protection are out of scope,
and we claim neither masking nor constant-time operation. On the prototype the wrapping key derives from the readable device ID, so a blob is tied to its device but not yet secret from its host. The 2 custody flaws (E2b) are circled in
Fig.~\ref{fig:custody}. The random generator (DRBG) and the key wrap could use the
\emph{shared hash core} at the same time, and the wrap took a \emph{host-chosen nonce}. Either custody flaw let such a
host recover a wrapped key, with all outputs correct. The 2 verifier flaws (both in row E2) are
the \emph{missing hint checks} and the \emph{unchecked $z$ coefficients} of a signature, which only malformed
signatures expose.

\begin{table}[t]
\caption{Review replays: sessions that found each flaw.}
\label{tab:replay}
\centering\scriptsize
\setlength{\tabcolsep}{2.5pt}
\begin{tabularx}{\columnwidth}{@{}l>{\raggedright\arraybackslash}Xrrr@{}}
\toprule
Code given, scope & Flaw & Plain request & Lens prompt & All \\
\midrule
A, part named$^{a}$ & each known flaw & 9/9 & 9/9 & 27/27 \\
A, whole design & shared hash core & 6/6 & 5/6 & 11/12 \\
 & host-chosen nonce & 6/6 & 6/6 & 12/12 \\
 & missing hint checks & 3/6$^{b}$ & 6/6 & 9/12 \\
 & unchecked $z$ coefficients & 1/6$^{b}$ & 6/6 & 7/12 \\
B, part named & planted, custody$^{c}$ & 1/8 & 2/12 & 3/20 \\
 & planted, verifier$^{c}$ & 8/8 & 12/12 & 20/20 \\
\midrule
 & none: false alarms$^{d}$ & 2/12 & 1/12 & 3/24 \\
\bottomrule
\end{tabularx}\\[2pt]
\parbox{\columnwidth}{\scriptsize $k/n$: sessions that found the flaw (last row: raised a false alarm). A
flaw counts as found when two judge agents (a third on ties), not told the prompt, label a finding as describing it
in the right file and logic.
$^{a}$First replay: 54 sessions = 2 parts $\times$ 3 prompts $\times$ 9, median 6 minutes; its registered
predictions: 4 wins, 8 ties at the ceiling; All adds a third prompt
(plain request plus threat model and design claims; also 9/9). The judges labelled 758 findings
($\kappa\ge0.97$ on the four flaws); auditor agents checked all 108 matches against the code (2 state the flaw with a
false side claim). Other rows: second replay, same session set-up (planted and clean sessions: only B's RTL), 76 sessions = 12 whole design (the same 12 for all four flaws) + 40
planted + 24 clean; one re-run because it had read a saved copy of its own tool output outside its directory; judges
disagreed on 2 of 1,966 items; matches not re-checked against the code. Of the lead agent's 15 range predictions for
this replay, 5 hit and 10 missed.
$^{b}$Against 9/9 in the first replay: Fisher $p=0.044$ and $0.002$ (descriptive). $^{c}$8 planted flaws, 5
sessions each (2 plain, 3 lens): custody flaws found in 0, 1, 0 and 2 of 5 (second and third: their formal check
is a nonce assertion added for this replay), each verifier flaw in 5 of 5. $^{d}$Clean
code, 12 sessions per part; all 3 alarms in the custody part; no clean session was credited with a planted flaw
(0/24), a control on the judges.}
\end{table}

\textbf{Review pointed at flawed code in stage 5.} All 54 \emph{sessions} of the first \emph{replay}, a new review
of A's unchanged RTL, found every known flaw in the \emph{part} they were pointed at (Table~\ref{tab:replay}).
Each session, a fresh
read-only run of a reviewer agent (claude-opus-5-5), worked in a new directory holding only A's RTL, with
file-reading tools and none of the project's instructions, memory, records or B code. Its prompt named the files of
one part: the custody
path or the verifier. A \emph{plain request} (no lens, no claims) did as well as the \emph{lens prompt} (with both), the flow's
reviewer template adapted to read-only review. So where a part is named, the prompt shows no gain;
what the flow adds is scope (stage~5 reviews one unit, not the whole design) and obligation (a review before each commit). We chose the parts knowing the flaws, so this replay shows that a review
finds them once it looks there, not how often it looks. The flaws stayed hidden for the port's first 7 to 13 weeks, until agents of this model found them (Sec.~\ref{sec:threats}).

\textbf{Review of a whole design in stage 5.} Given all of A and a plain request, reviewers found both custody flaws
but often missed the verifier flaws (Table~\ref{tab:replay}). The second replay gave 12 sessions all of A,
6 per prompt, naming no part. Only 3 of 6 plain-request sessions found the missing hint checks and 1 of 6 the
unchecked $z$ coefficients. The lens prompt found each in 6 of 6, but its design claims point to both
flawed parts, so the plain request was pre-registered as the primary measure.

\textbf{Planted flaws in stage 5.} The first replay's flaws were ones the model had found before, so a seeded
script planted new ones in B. A \emph{planted flaw} is one random one-line edit (a term deleted, a comparison
negated or a constant off by one) in code of a known flaw class, kept only if the functional checks pass and a security check
fails: a formal check (custody) or a reject test (verifier). Three of the four custody ones also fail the custody
testbench's security checks. Pointed at the flawed part, all 20 verifier sessions but only 3 of 20 custody sessions
found it (Table~\ref{tab:replay}). Possible reasons,
none tested: A's custody flaws are missing mechanisms, while a planted
one changes one term; and it sits 13--59 lines from the comment stating its invariant, while 3 of 4 planted
verifier flaws sit next to one.

\textbf{Clean code in stage 5.} On clean code, 3 of 24 sessions raised a \emph{false alarm}, a claimed flaw of a known
class (Table~\ref{tab:replay}), all in the custody part. All 3 describe one problem in how the hash core changes hands, real within the
reviewed RTL but prevented by a wrapper outside it. A post-hoc simulation with a negative control confirmed this for
key generation, and code reading for the wrap, so the review's scope caused them.

\textbf{Formal checks of the custody flaws.} Open-source formal tools (SymbiYosys with ABC and rIC3), with no LLM as
judge, checked the custody flaws on an abstraction and properties that one agent wrote. It keeps the
arbitration and custody logic as written and leaves the engines, memories and every host input unconstrained. This
only adds behaviours, so a proof covers the core's RTL. The property is that the hash core never has two owners, not
that no secret leaks. On A, a bounded check found the random generator and the key wrap holding the core together in the
third cycle after reset, once two inputs driven by host-written registers had changed. Simulating the unabstracted
core reproduced it. On B two engines proved the property without a bound. With one exclusion term removed, the proof
failed, as it should. The proof does not rely on the wrapper that prevents the clean-code alarms, and says nothing about the
core's state when it changes hands. A static trace of every signal feeding the wrap
nonce found a host register on A and only the random generator on B, except when migration re-wraps a key of the old format. We did not check that case.

\section{Limits and Threats to Validity}
\label{sec:threats}

\textbf{Semi-automated, one design family.} The operator ruled on scope and questions and performed every
board action; the lead agent launched stage~5's rounds. The study covers one design family: A was built under the protocol of~\cite{park2026aiassisted},
and B was ported from A's sources under the full workflow. Public code for both schemes may be in
the training data~\cite{wang2025vericontaminated}.

\textbf{One operator, no baseline, self-reported telemetry.} One experienced operator
worked with one model family, and every replay session used one model; there is no human-only or open-model baseline~\cite{baltes2026guidelines}, hence no
speed-up claim. Agents of the replays' model first found all four known flaws during the port, so the replays cannot
count flaws this model misses. None of the four was found before that model's first use. Agents wrote the logs at close-out, operator minutes and task category included, knowing the outcome; a unit is
logged once, so retries become iterations, not failures, and the rate partly measures what counts as a unit.

\textbf{Evidence limits.} Each board session ran once, on one board: the known answers and ACVP cases passed
again in later sessions; the performance figures were not repeated. A safety rule bars sending the device
malformed signatures, so the verifier's added checks are shown only in simulation (boundary
testbenches; 14 mutants, all killed). The raw entropy source has no min-entropy estimate. Only late sessions have
predictions: in six graded ones \cTwoPredMet{} of \cTwoPredRegistered{} held (\cTwoPredHashRegistered{} hashed beforehand); later
boot-chain sessions are not counted. We froze the plan in our own repository, not a registry; all telemetry comes
from one snapshot of work in progress. Planted flaws are artificial (random edits in regions chosen knowing A's
flaws), and their class cannot be told apart from comment distance. Replay judges share the reviewers' model and
knew each planted line. Neither mutant-experiment agent was blind: both had read the predictions, and the test
writer knew the fresh draw's seed; by its log it drew only after the freeze.

\section{Conclusion}

Passing tests and approving reviews from one workflow are not enough to trust agent-built hardware. On an
ML-KEM-768/ML-DSA-65 port that passes all 85 interface-applicable NIST ACVP cases, the passing simulation gates of
three modules kill only 76\% of tool-selected mutants; undirected review often missed the verifier flaws, and scoped
review found planted custody flaws in only 3/20 sessions, while formal checks (one added for the replay) flag every
one. In this study, trustworthy agentic hardware needed author-independent mutants for test gates, module-level
scoping for LLM review, and formal checks for custody invariants. Next: cross-model reviewers and from-scratch design
on other hardware families.

\section*{Content Generated by AI}

LLM agents in the Claude Code harness did the work studied here under the operator's direction. The logs name claude-opus-5-5 (\cTwoModelOpusFiveFiveN{} units), claude-fable-5 (\cTwoModelFableFiveN), claude-fable-5-1
(\cTwoModelFableFiveOneN), claude-fable-5-2 (\cTwoModelFableFiveTwoN), claude-opus-5 (\cTwoModelOpusFiveN) and claude-opus-4-8 (\cTwoModelOpusFourEightN); the token ledger adds \cTwoTokModelsNotInLogsIds{} (under 0.1\%). Agents
of these models drafted all sections' text, tables, figures and statistics scripts under the authors' direction; the authors revised all of it and take full responsibility for it.

\clearpage
\bibliographystyle{IEEEtran}
\bibliography{refs}

\end{document}